# *Multiferroic properties of o-LuMnO$_3$ controlled by b-axis strain*

Y. W. Windsor[1], S. W. Huang[1], Y. Hu[2], L. Rettig[1], A. Alberca[1], K. Shimamoto[2], V. Scagnoli[1] T. Lippert[2], C. W. Schneider[2], U. Staub[1]

[1]*Swiss Light Source, Paul Scherrer Institut, 5232 Villigen PSI, Switzerland*

[2]*General Energy Research Department, Paul Scherrer Institut, 5232 Villigen PSI, Switzerland*

Strain is a leading candidate for controlling magnetoelectric coupling in multiferroics. Here, we use x-ray diffraction to study the coupling between magnetic order and structural distortion in epitaxial films of the orthorhombic (o-) perovskite LuMnO$_3$. An antiferromagnetic spin canting in the E-type magnetic structure is shown to be related to the ferroelectrically induced structural distortion and to a change in the magnetic propagation vector. By comparing films of different orientations and thicknesses, these quantities are found to be controlled by *b*-axis strain. It is shown that compressive strain destabilizes the commensurate E-type structure and reduces its accompanying ferroelectric distortion.

Recent years have seen significant efforts in the study of multiferroic and magnetoelectric materials, which exhibit both ferroelectric and magnetic orders simultaneously. These materials are important due to the prospect of controlling magnetization (electric polarization) with electric (magnetic) fields in future technological applications. This might be achieved in single phase materials with strong magnetoelectric couplings [1–3]. Manganites with an o-*RE*MnO$_3$ orthorhombic perovskite structure (*RE* denoting a heavy rare earth) are seen as prototypical multiferroic materials. Within the o-*RE*MnO$_3$ family, materials with *RE* equal to- or heavier than Ho have been shown to possess an E-type antiferromagnetic (AFM) order (↑↑↓↓ type in the *ab* plane) at low temperatures [4–7]. It has been suggested that this E-type spin structure may induce a substantially large electric polarization *P* of up to 120 mC/m$^2$, due to symmetric exchange striction [8,9]. Indeed polycrystaline o-*RE*MnO$_3$ with *RE*=Y, Ho, Tm, Yb and Lu have been shown to exhibit relatively large values of up to 0.8 mC/m$^2$, which may correspond to nearly 5 mC/m$^2$ in a single crystal [10], and values up to 8 mC/m$^2$ have been reported for epitaxial films with *RE*=Y [11]. In contrast, ferroelectricity is weaker in o-*RE*MnO$_3$ materials with lighter *RE* ions (e.g. *RE*=Tb), where it is caused by a cycloidal antiferromagnetic order [12]. Unfortunately, bulk o-*RE*MnO$_3$ samples with heavier *RE* ions can only be synthesized under high oxygen pressure [10], significantly limiting studies on these interesting materials due to the absence of large high-quality single crystals.

*RE*=Lu has the highest ferroelectric transition temperature $T_C$ of the o-*RE*MnO$_3$ series [10]. At room temperature its structure is described by the Pbnm space group. It posesses a sinusoidal antiferromagnetic order below $T_N$≈42K and an E-type antiferromagnetic order below the ferroelectric transition at $T_C$≈35K [10]. Large polarizations have been reported for the latter phase, but previous works focused only on powders and polycrystalline samples. Success in studying epitaxial films of o-LuMnO$_3$ has recently been reported [13], indicating that high quality functional samples can be synthesized. Furthermore, the prospect of manipulating magnetoelectric properties through strain introduces a new degree of freedom for controlling functionality in future spintronic applications. It is therefore of great interest to study epitaxial and single crystalline samples of this promising magnetoelectric material.

Resonant soft x-ray diffraction (RSXD) [14] has been used in recent years to study the magnetic ordering in multiferroic single crystals of o-REMnO$_3$ with e.g. RE=Tb [15,16] and RE=Dy [17], and in films of RE=Y [18]. RSXD is element-selective, allowing one to directly probe the Mn 3d (magnetic) states due to a transition from the $2p_{1/2}$, $2p_{3/2}$ to the 3d states at the Mn $L_{2,3}$ absorbtion edges. This technique is especially suitable for studying magnetism in thin films (as demonstrated on NdNiO$_3$ [19]) because even small sample volumes can be used due to the large resonant enhancement of magnetic scattering at the transition-metal 2p→3d absorption edges.

In this work we study the magnetic ordering of epitaxially grown films of o-LuMnO$_3$ using RSXD at the Mn 2p→3d absorption edge. We find that the films notably differ from known bulk behavior in that the magnetic diffraction peak's position varies significantly with temperature (T), in contrast to studies on bulk o-REMnO$_3$ materials with E-type order. The evolution of the structural distortion caused by the electric polarization is probed using non-resonant x-ray diffraction and is found to be directly related to changes in the magnetic order. Epitaxial strain is shown to control them both.

The films have orthorhombic structure at room temperature, and were epitaxially grown to thicknesses ranging from 26 to 200 nm, on [010]- and [110]-oriented YAlO$_3$ single crystalline substrates [20], thus varying the effective strain. The films were prepared by pulsed laser deposition (PLD) using a KrF excimer laser ($\lambda$ = 248 nm, 2 Hz) with a laser fluence of 3 J/cm$^2$ and a substrate temperature 760°C. Stoichiometric sintered ceramic targets of hexagonal LuMnO$_3$ were used, which were prepared by conventional solid state synthesis. To provide more atomic oxygen for the film growth, N$_2$O was used as background gas ($p$N$_2$O=0.3 mbar). By using Cu $K_\alpha$ x-ray diffraction, the films were shown to be single-phase, untwinned and of single crystalline quality [21].

RSXD experiments were carried out using the RESOXS station [22] at the X11MA beamline of the Swiss Light Source [23], at the Mn $L_{2,3}$ edges ($h\nu$=652.2 and 643 eV). The RSXD experimental geometry is the same as Ref. [18]. The azimuthal angle $\Psi$ (angle of sample rotation around the momentum transfer vector $\vec{q}$) is 0 when the *c*-axis is perpendicular to the scattering plane. Hard x-ray diffraction (XRD) experiments were performed at the Surface Diffraction endstation of the X04SA beamline at the Swiss Light Source [24] with $h\nu$=8 KeV, and the lattice parameters were determined for temperatures down to 19K using a He flow cryostat.

Magnetism of the Mn ions is studied through the RSXD intensity of the magnetic (0 $q_b$ 0) reflection with $q_b \approx \frac{1}{2}$. To study the spin structure below $T_C$ [10], the intensity's azimuthal angle dependence was recorded at 10K using linearly polarized incident light. This is presented in the inset of Fig. 1. At $\Psi$=0° the intensity is equal for π and σ linear polarizations (E-field in and perpendicular to the scattering plane, respectively); at other angles the intensity for π is larger than for σ. The difference is maximal at $\Psi$=90°. Calculating the magnetic structure factor $F = (\varepsilon \times \varepsilon') \sum_i \hat{m}_i \exp(2\pi i \vec{q} \cdot \vec{r}_i)$ [25] of an E-type magnetic structure indicates that the intensity of the (0 ~½ 0) reflection is sensitive only to $S_{\hat{c}}$, the Mn spin projections along the *c*-axis, and not to the main *b*-axis spin component ($\varepsilon$ and $\varepsilon'$ are the incoming and outgoing polarizations, the sum is over all magnetic ions in the supercell, and $\hat{m}_i$ and $r_i$ are the moment directions and positions of the *i*-th ion, respectively). This suggests the existence of an E-type structure with an additional alternating spin-canting component along this axis ($S_{\hat{c}}$). Such canting was previously reported in powder samples of o-LuMnO$_3$ and epitaxial o-YMnO$_3$ films [18,26], and has been theoretically predicted [27]. The solid lines in the inset of Fig. 1 are calculations of the expected intensities for this case, and are in excellent agreement with the experimental data (circles).

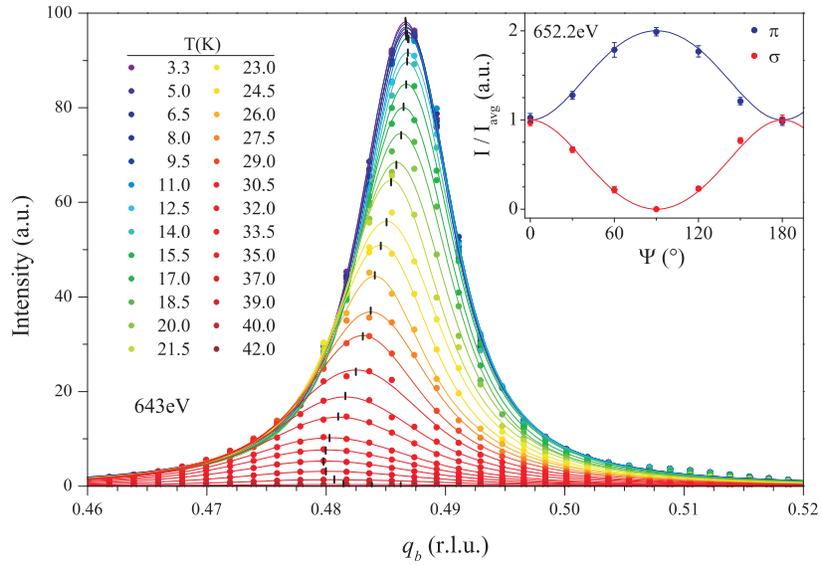

Fig. 1: Temperature dependence of the magnetic (0 $q_b$ 0) reflection with $q_b \approx ½$ of a 104nm [010]-oriented film of o-LuMnO$_3$ ($h\nu$=643eV). The fitted peak positions are marked by ticks. Inset: Azimuthal angle $\Psi$ dependence of the (0 $q_b$ 0) intensity in o-LuMnO$_3$ film for π and σ polarized incident light (normalized by the average intensity of π and σ), measured at 10K and $h\nu$=652.2 eV. Lines represent a calculation of the expected intensity in an E-type AFM with spin canting along the *c*-axis.

The temperature dependence of this reflection in a 104 nm [010]-oriented film is presented in Fig. 1 for incident π polarized light. The reflection appears below the Néel temperature $T_N \approx 42K$, and its intensity sharply increases with the onset of E-type order [26], down to saturation at ~11K, as shown in inset (a) of Fig. 2. The peak position deviates from $q_b$=0.5 at all measured temperatures, indicating an incommensurability (ICM) of the magnetic structure. Furthermore, $q_b$ is clearly temperature dependent above a lock-in temperature $T_L \approx 14K$ (see marked peak positions in Fig. 1). This striking feature is not related to lattice contractions [28] and has not been reported for bulk samples of o-LuMnO$_3$, where the magnetic structure is expected to be a fully-commensurate AFM E-type [7]. A similar observation was recently reported for o-YMnO$_3$ thin films [18], which also belong to the o-*RE*MnO$_3$ family with E-type order. Unlike the intensity, $q_b$ does not evolve monotonically with *T*, but changes direction at *T*\*≈37K. It is possible that for *T*\*<*T*<*T*$_N$ the canted E-type and sinusoidal AFM phases coexist, resulting in a weak and possibly indistinguishable second peak. Coexistence with a spin-cycloid, as found for a large temperature range in o-YMnO$_3$ [18], cannot be excluded either as again it may manifest as a weak peak with a very similar wave vector. Indeed nonzero spin helicity vectors have been predicted in the *a* and *c* directions [27]. Studies on other bulk o-*RE*MnO$_3$ systems with an E-type order have also reported a constant $q_b$ below $T_C$, in contrast to our thin film results, and the change in trend of $q_b(T)$ at *T*\* was also absent [4,5,29]. Fig. 2 shows the *T*-dependence of deviation from $q_b(T_L)$, defined as $\Delta q_b(T) = q_b(T) - q_b(T_L)$, for a series of films. $q_b(T_L)$ itself is always very close to 0.5, but varies slightly between samples. This is consistent with Monte Carlo simulations that predict deep minima on the energy landscape of the ordering wavevector near (0 ½ 0) [27].

A second observation was made in the same *T* range: a phenomenological relation exists between the integrated intensity $I(T)$ and the magnetic modulation $q_b(T)$. The relation follows $I(T) = \exp(\alpha q_b(T) + \beta)$, holds for $T_L$<*T*<*T*\*, and is unequivocally satisfied in all tested films. While the origin of this relation is not understood, its importance stems from the fact that it is always satisfied, regardless of where our samples fall on the energy landscape (indicated by $q_b(T_L) \neq 0.5$ values). Since $I(T) \propto |F|^2$, the parameter *β* is a measure of the magnitude of Mn spin canting along the *c*-axis ($S_{\hat{c}}$) [30], while *α* is a measure of how $S_{\hat{c}}$ relates to the incommensurability of magnetic modulation, and is a positive number. These parameters exhibit a linear correlation, indicating that the incommensurate modulation is directly related to the magnitude of $S_{\hat{c}}$.

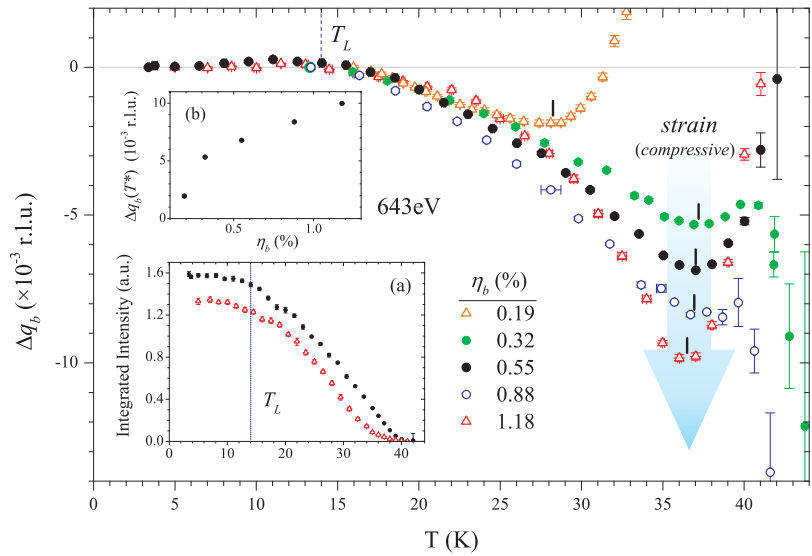

Fig. 2: Temperature dependences of fitted parameters from the (0 $q_b \approx \frac{1}{2}$ 0) reflection for various films labeled by $\eta_b$, their *b*-axis strain at low *T*. Main figure: change in modulation $\Delta q_b$ from the locked-in state $q_b \approx \frac{1}{2}$ at $T_L$. $T^*$ values are marked by ticks. Inset (a): corresponding dependence of the integrated intensity for selected films. Inset (b): variation of $\Delta q_b$ ($T^*$) with *b*-axis strain. Samples: triangles indicate [110]-oriented films ($\eta_b$ =0.19 and 1.18% correspond to 200 and 90 nm thick films, respectively), circles indicate [010]-oriented films ($\eta_b$=0.32, 0.55 and 0.88% correspond to 78, 104 and 26 nm thick films, respectively).

The onset of E-type AFM order creates a spontaneous ferroelectric polarization at $T_C$ [10], which is associated with a lowering of crystal symmetry from Pbnm. Hard x-ray diffraction was used to probe Pbnm-forbidden structural reflections (0 $k$ 0) with $k$-odd, that are sensitive to the corresponding structural distortion for $T<T_C$. Fig. 3 shows the temperature dependence of the (0 5 0) reflection of a 200nm [110]-oriented film. The reflection appears below $T_C \approx 40K$, and is directly sensitive to the ionic shifts in the $b$-direction, $\delta_b$, which are expected to be proportional to the structural distortion (ferroelectric polarization). The integrated intensities of this reflection are presented in inset (a) of Fig. 3, alongside those of the magnetic (0 $q_b$ 0) reflection from this film. The similar shape and onset of the $T$-dependence indicates that the ionic shifts $\delta_b$ and the Mn spins' c-axis component $S_{\hat{c}}$ arise from- and are correlated to the same order parameter, the one associated with the E-type magnetic order.

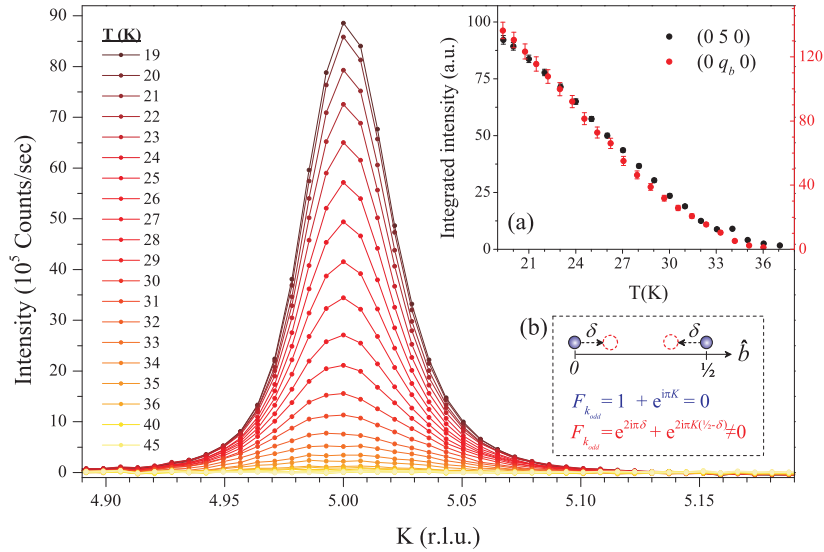

Fig. 3: Temperature dependence of the (0 5 0) reflection from a 200nm o-LuMnO$_3$ [110]-oriented film. Inset (a): Temperature dependence of the integrated intensity from the (0 5 0) structural reflection (black) and the (0 $q_b \approx$½ 0) magnetic reflection (red) of this film. Inset (b): simplest approximation for a distortion producing nonzero intensity for a (0 $k$ 0) reflection with $k$-odd, depicted for 2 atoms along the $b$-direction.

The (0 $k$ 0) reflections with $k$ odd were several orders of magnitude weaker than the Pbnm-allowed (0 2 0) reflection, indicating that the structural distortion is weak. The exact structural distortion is unknown, but is expected to be of the same type in all films. We therefore use the simplest approximation for the structure factor of such weak reflections to compare films (see inset (b) in Fig. 3). This can be expressed as $F(k,E) \approx \bar{f}(k,E) \cdot \sin(2\pi \cdot k \cdot \bar{\delta}_b) \approx 2\pi \cdot \bar{f}(k,E) \cdot k \cdot \bar{\delta}_b$, in which $\bar{f}(k,E)$ is the averaged form factor of all ions, and $\bar{\delta}_b$ is the average ionic shift within this simplistic model. $\bar{\delta}_b$ can be obtained by normalizing the calculated and observed intensities from (0 $k$ 0) with $k$ odd to those of the (0 2 0) reflections. Intensities are corrected for scattering volume and polarization factors. The $T$-dependence of $\bar{\delta}_b$ is presented in Fig. 4 for different films. Within the approximation of a small distortion, we expect $\bar{\delta}_b$ from this model to produce reasonable approximations of the actual distortion, so the size of $\bar{\delta}_b$ is assumed to be proportional to that of the ferroelectric polarization. Validity of the approximation can be checked by comparison with the full crystal structure below $T_C$. This requires collecting a full crystallographic data set from a single crystal, which exists only for o-$RE$MnO$_3$ with $RE$=Y [31]. Adopting the reported structures above and below $T_C$ [32], we can quantitatively determine the distortion of the ions [33]. For example, in an $\eta_b$=0.19% strained film, the largest displacement of Mn ions along the $a$ axis is 0.0015 Å, compared to 0.0041Å for the o-YMnO$_3$ crystal in Ref. [31] (both at 21K). Values found using the structures reported for $RE$=Y are within 25% of $\bar{\delta}_b$, indicating that this simplistic approximation is adequate, and $\bar{\delta}_b$ therefore serves as a measure to compare the effective ferroelectric distortions of different films.

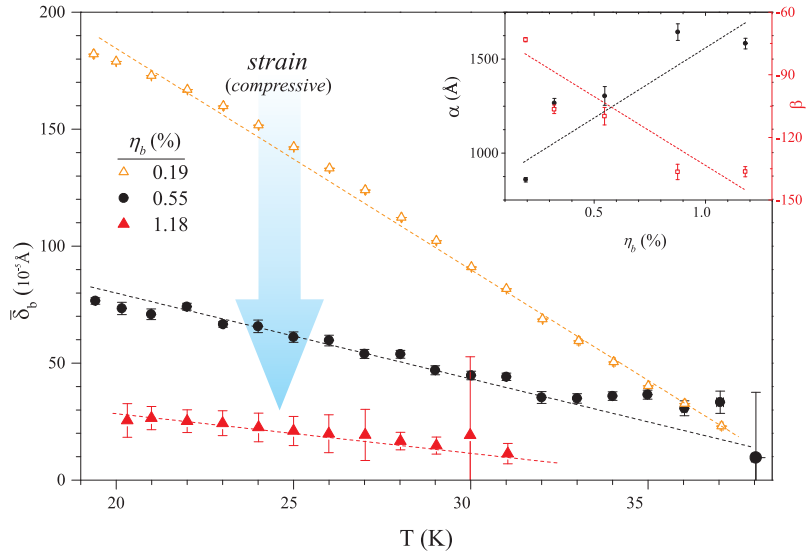

Fig. 4: Temperature dependence of the average ionic shift in the b-direction $\bar{\delta}_b$ for different films labeled by their $\eta_b$ values. Samples: triangles indicate [110]-oriented films ($\eta_b$ =0.19 and 1.18% correspond to 200 and 90 nm thick films, respectively), circles indicate a [010]-oriented 104 nm thick film. Inset: $\alpha$ and $\beta$ parameters as functions of b-axis (compressive) strain. Lines are guides for the eyes.

The strain in each film can be estimated by comparing measured lattice constants to those of bulk [34], as $\eta_x=(x_{bulk}-x_{film})/x_{bulk}$ ($x$ represents a lattice constant). Since film thickness and substrate orientation both affect strain, we discuss strain observed along each crystal direction independently [35]. We find that only the b-axis (compressive) strain $\eta_b$ plays a clear role. The value of $\bar{\delta}_b$ (at 20K) grows significantly as $\eta_b$ is reduced (see Fig. 4). Indeed it has been recently predicted that reducing compressive ab-strain increases the contribution to P arising from ionic displacements [36]. $S_{\hat{c}}$, followed through the $\beta$ parameter, also grows as $\eta_b$ is reduced. However, $\alpha$ (the ICM contribution) is enhanced by larger values of $\eta_b$ (see inset of Fig. 4), indicating that the canted E-type structure is destabilized by compressive b-axis strain. This is supported by the trend in the size of $\Delta q_b(T^*)$ which increases approximately linearly with $\eta_b$ (inset (b) of Fig. 2). Since $S_{\hat{c}}$ is coupled to the same order parameter as $\bar{\delta}_b$, it is linear to P. Variations in $S_{\hat{c}}$ are therefore not expected to contribute to P through symmetric exchange striction, which relates the squared spin size to electric polarization P by $P \propto \sum \vec{S}_i \cdot \vec{S}_j$ [27]. Curiously, the c-axis strain remains at ~ -1% in all films, and does not appear to relax, whereas the a-axis strain varies but remains small. We note that a clear strain-dependent trend was not identified for $T_L$, $T^*$ and $T_C$, for example we find that 33K<$T_C$<40K while $T_L$ appears constant for all films, which might be correlated to the smaller lattice strain in the a-direction. Our findings indicate that the magnetic and electric properties are controlled by b-axis strain. This is supported by theoretical calculations, that indicate that $J_b$, the exchange coupling along the b-axis, is the crucial parameter that determines the magnetic ground state [27].

In summary, we have studied the magnetic order and ferroelectric distortion of epitaxial o-LuMnO$_3$ films using resonant soft- and non-resonant hard x-ray diffraction. The azimuthal-angle dependence of the (0 ~½ 0) magnetic reflection is consistent with a c-axis magnetic moment contribution to E-type AFM order, which grows significantly below $T_C$. The magnetic modulation vector is incommensurate and temperature dependent. The temperature dependence of Pbnm-forbidden structural reflections was found to be similar to that of the magnetic (0 ~½ 0) reflection, indicating that the ionic shift and the c-axis spin component arise from the same order parameter. Most importantly, it is found that b-axis compressive strain weakens both quantities, while the effect of magnetic incommensurability increases with strain. This is emphasized with a comparison to single crystal o-YMnO$_3$, where the structural distortion is 2.7 times larger than in our least-strained film. It is thus understood that strain strongly controls the stability of the commensurate E-type structure and its accompanying ferroelectric distortion. Such results are likely to appear also in multiferroic manganite films with other RE ions, and they agree with the theoretical prediction that tensile b-axis strain will significantly enhance the polarization in these systems arising from ionic displacements.

We gratefully thank the X11MA and X04SA beamline staff for experimental support. The financial support of PSI, the Swiss National Science Foundation and its NCCR MaNEP is gratefully acknowledged.

**References**

[1]  Y. Tokura, Science **312**, 1481 (2006)

[2]  S.-W. Cheong et al., Nature Mater. **6**, 13(2007)

[3]  Y. Tokura et al., Adv. Mater. **22**, 1554 (2010)

[4]  A. Muñoz et al., Inorg. Chem. **40**, 1020 (2001)

[5]  V.Yu. Pomjakushin et al., New J. Phys. **11**, 043019 (2009)

[6]  M. Tachibana et al., Phys. Rev B **75**, 144425 (2007)

[7]  H. Okamoto et al., Solid State Commun. **146**, 152 (2008)

[8] I. A. Sergienko et al., Phys. Rev. Lett. 97 227204 (2006)

[9]  I. A. Sergienko et al., Phys. Rev B **73**, 094434 (2006)

[10] S. Ishiwata et al., Phys. Rev B **81**, 100411(R) (2010)

[11] M. Nakamura et al., Appl. Phys. Lett. **98**, 082902 (2011)

[12] M. Mostovoy, Phys. Rev. Lett. **96**, 067601 (2006)

[13] J.S. White et al., Phys. Rev. Lett. **111**, 037201 (2013)

[14] J. Fink et al., Rep. Prog. Phys. **76,** 056502  (2013)

[15] S.B. Wilkins et al., Phys. Rev. Lett. **103**, 207602 (2009)

[16] H. Jang et al., Phys. Rev. Lett. **106**, 047203 (2011)

[17] E. Schierle, Phys. Rev. Lett. **105**, 167207 (2010)

[18] H. Wadati et al., Phys. Rev. Lett. **108**,047203 (2012)

[19] V. Scagnoli et al., Phys. Rev. B **73**, 100409(R) (2006)

[20] Substrates from Crystec GmbH, Berlin,

[21] Y. Hu et al., Appl. Phys. Lett. **100**, 252901 (2012)

[22] U. Staub et al., J. Synchrotron Radiat. **15**, 469 (2008)

[23] U. Fleschig et al., AIP Conf. Proc. 1234, 319 (2010)

[24] P. R Willmott et al., J. Synchrotron Rad. **20**, 667–682 (2013)

[25] J. P. Hill and D. F. McMorrow, Acta Cryst. **A52**, 236-244 (1996)

[26] M. Garganourakis et al., Phys. Rev. B **86**, 054425 (2012)

[27] M. Mochizuki et al., Phys. Rev. B **84**, 144409 (2011)

[28] Changes with T in the (020) structural reflection were found to be 100 times smaller than those in (0 $q_b$ 0).

[29] F. Ye et al., Phys. Rev. B **76**, 060402(R) (2007)

[30] $\beta$ is related to the angle of Mn spin canting in the bc-plane. This angle cannot be decoupled from the ordered moduli of the magnetic moment as (0 ½ 0) is the only magnetic reflection obtainable at the Mn *L* edges.

[31] D. Okuyama et al., Phys. Rev. B **84**, 054440 (2011)

[32] The great similarity in the ferroelectric behavior for *RE*=Lu and Y permits using the structures reported for *RE*=Y above and below $T_C$, with the lattice parameters found for *RE*=Lu.

[33] To do so we calculate how the ratio of intensities $I_{(0k0)}/I_{(020)}$ (with *k* odd) changes from the structures reported above and below $T_C$. By comparing this to the measured values of $I_{(0k0)}/I_{(020)}$, we deduce the shifts of all ions in the unit cell. We note that in essense $I_{(0k0)}/I_{(020)}$ *is* the same quantity used to calculate $\bar{\delta}_b$, only here phase contributions to the structure factor are taken into account.

[34] H. Okamoto et al., Sol. Stat. Comm. 146 (2008) 152–156

[35] $\eta_b$ values for 3 films were estimated from room temperature data, normalized by the contraction of a similar film down to 20K.

[36] D. Iuşan et al., Phys. Rev. B **87**, 014403 (2013)